\documentclass[final]{siamltex}

\usepackage{epsfig,amssymb,latexsym}
\usepackage{amsfonts,psfrag,amsmath,bbm,color}
\usepackage{cancel}

\usepackage{mathrsfs}
\usepackage{graphicx}
\usepackage{textcomp}
\usepackage{multirow}
\usepackage{enumerate}
\usepackage{cancel}
\usepackage{caption}  
\usepackage{url}

\usepackage{algorithm}
\usepackage{algorithmic}

\usepackage{amscd}

\newtheorem{remark}{Remark}[section]

\def\PP{{{\rm l}\kern - .15em {\rm P} }}
\def\PN2{{\PP_{N}-\PP_{N-2}}}

\newcommand{\R}{\mathbbm{R}}

\newcommand{\cO}{\mathcal{O}}

\newcommand{\bphi}{\boldsymbol{\varphi}}

\newcommand{\btau}{\boldsymbol{\tau}}

\newcommand{\ba}{\boldsymbol{a}}
\newcommand{\bas}{{\boldsymbol a}^{snap}}

\newcommand{\bff}{\boldsymbol{f}}

\newcommand{\bu}{\boldsymbol{u}}

\newcommand{\bur}{{\boldsymbol{u}}_r}

\newcommand{\bx}{\boldsymbol{x}}
\newcommand{\bX}{\boldsymbol{X}}

\newcommand{\bXr}{{\bf X}^r}

\newcommand{\tA}{\tilde{A}}
\newcommand{\tB}{\widetilde{B}}

\newcommand{\obu}{\overline{\boldsymbol u}}

\newcommand{\as}{a^{snap}}

\newcommand{\deleted}[1]{{}}

\begin{document}

\title{Data-Driven Filtered Reduced Order Modeling \\ Of Fluid Flows}

\author{
X. Xie
\thanks{Department of Mathematics, Virginia Tech, Blacksburg, VA 24061
Partially supported by NSF DMS1522656, email: xupingxy@vt.edu}
\and M. Mohebujjaman
\thanks{Department of Mathematical Sciences, Clemson University, Clemson, SC, 29634;
Partially supported by NSF DMS1522191, email: mmohebu@g.clemson.edu}
\and L. G. Rebholz
\thanks{Department of Mathematical Sciences, Clemson University, Clemson, SC 29634; 
Partially supported by NSF DMS1522191 and Army Research Office 65294-MA, email: rebholz@clemson.edu}
\and T. Iliescu
\thanks{Department of Mathematics, Virginia Tech, Blacksburg, VA 24061
Partially supported by NSF DMS1522656, email: iliescu@vt.edu}.
}

\date{\today}

\maketitle

\begin{abstract}
We propose a data-driven filtered reduced order model (DDF-ROM) framework for the numerical simulation of fluid flows.
The novel DDF-ROM framework consists of two steps:
(i) In the first step, we use explicit ROM spatial filtering of the nonlinear PDE to construct a filtered ROM.
This filtered ROM is low-dimensional, but is not closed (because of the nonlinearity in the given PDE).
(ii) In the second step, we use data-driven modeling to close the filtered ROM, i.e., to model the interaction between the resolved and unresolved modes.
To this end, we use a quadratic ansatz to model this interaction and close the filtered ROM.
To find the new coefficients in the closed filtered ROM, we solve an optimization problem that minimizes the difference between the full order model data and our ansatz.  
We emphasize that the new DDF-ROM is built on general ideas of spatial filtering and optimization and is  independent of (restrictive) phenomenological arguments.

We investigate the DDF-ROM in the numerical simulation of a 2D channel flow past a circular cylinder at Reynolds number $Re=100$.
The DDF-ROM is significantly more accurate than the standard projection ROM.
Furthermore, the computational costs of the DDF-ROM and the standard projection ROM are similar, both costs being orders of magnitude lower than the computational cost of the full order model. 
We also compare the new DDF-ROM with modern ROM closure models in the numerical simulation of the 1D Burgers equation.
The DDF-ROM is more accurate and significantly more efficient than these ROM closure models.
\end{abstract}




\pagestyle{myheadings}
\thispagestyle{plain}
\markboth{X. XIE, M. MOHEBUJJAMAN, L. G. REBHOLZ, AND T. ILIESCU}{DATA-DRIVEN FILTERED REDUCED ORDER MODELING OF FLUID FLOWS}

\section{Introduction}
    \label{sec:introduction}

Reduced order models (ROMs) have been successfully used to reduce the computational cost of scientific and engineering applications that are governed by relatively few recurrent dominant spatial structures~\cite{antoulas2005approximation,ballarin2016fast,benner2015survey,bistrian2015improved,gunzburger2017ensemble,hesthaven2015certified,HLB96,noack2011reduced,perotto2017higamod,quarteroni2015reduced,stefanescu2015pod}.

One of the most popular classes of ROMs is the {\it projection ROMs (Proj-ROMs)}.
For a given general partial differential equation (PDE), the Proj-ROM strategy for approximating the PDE's solution $\bu$, is straightforward:
(i) Choose modes $\{ \bphi_1, \ldots, \bphi_d \}$, which represent the recurrent spatial structures of the given PDE.
(ii) Choose the dominant modes $\{ \bphi_1, \ldots, \bphi_r \}$, $r \leq d$, as  basis functions for the ROM.
(iii) Use a Galerkin truncation $\bur = \sum_{j=1}^{r} a_j \, \bphi_j$.
(iv) Replace $\bu$ with $\bur$ in the given PDE.
(v) Use a Galerkin projection of PDE($\bur$) onto the ROM space $\bXr := \text{span} \{ \bphi_1, \ldots, \bphi_r \}$ to obtain a low-dimensional dynamical system, which represents the Proj-ROM. 
For example, in fluid dynamics, the Proj-ROM often takes the following form:
\begin{eqnarray}
	\dot{\ba}
	= A \, \ba
	+ \ba^{\top} \, B \, \ba \, ,
	\label{eqn:proj-rom}
\end{eqnarray}
where $\ba$ is the vector of unknown ROM coefficients and $A \in \R^{r \times r}, B \in \R^{r \times r \times r}$ are ROM operators.
(vi) In an offline stage, compute the ROM operators. 
(vii) In an online stage, repeatedly use the Proj-ROM~\eqref{eqn:proj-rom} (for various parameter settings and/or longer time intervals).
The Proj-ROM~\eqref{eqn:proj-rom} is often efficient and relatively accurate~\cite{ballarin2015supremizer,HLB96,noack2011reduced}, but can fail in realistic applications
when large numbers of modes are needed to accurately represent the system. 
To ensure a low computational cost, it is desirable that Proj-ROMs use only a few modes (i.e., low $r$ values) and discard the remaining modes $\{ \bphi_{r+1}, \ldots, \bphi_d \}$.
The resulting Proj-ROM, however, can yield inaccurate results (see, e.g.,~\cite{amsallem2012stabilization,balajewicz2013low,barone2009stable,benosman2017learning,bergmann2009enablers,carlberg2017galerkin,giere2015supg,osth2014need,protas2015optimal}).
The general explanation for these inaccurate results is that the Proj-ROM~\eqref{eqn:proj-rom} fails to account for the interaction between resolved and unresolved modes~\cite{AHLS88,CSB03,gouasmi2016characterizing,noack2008finite,wang2012proper,wells2017evolve,xie2017approximate}.
Thus, in practical applications, the following modified Proj-ROM is used:
\begin{eqnarray}
	\dot{\ba}
	= A \, \ba
	+ \ba^{\top} \, B \, \ba
	+ \btau \, ,
	\label{eqn:proj-rom-tau}
\end{eqnarray}
where $\btau$ models the interaction between resolved modes $\{ \bphi_1, \ldots, \bphi_r \}$ and unresolved modes $\{ \bphi_{r+1}, \ldots, \bphi_d \}$.
Most often, a dissipation mechanism (e.g., eddy viscosity) is used to model $\btau$ in the modified Proj-ROM~\eqref{eqn:proj-rom-tau}: 
\begin{eqnarray}
	\btau
	\approx EV(\ba) \, .
	\label{eqn:proj-rom-tau-ev}
\end{eqnarray}

\bigskip

Another class of ROMs is the {\it data-driven ROMs (DD-ROMs)}, which are an extremely dynamic research area, and are fundamentally different from the Proj-ROMs presented above.
Although both the DD-ROM and the Proj-ROM can be written as in~\eqref{eqn:proj-rom}, the operators $A$ and $B$ are constructed using fundamentally different approaches:
the Proj-ROMs use the Galerkin projection (as explained above), whereas the DD-ROMs use the available full order model (FOM) or experimental data~\cite{cacuci2013computational,kutz2013data}.
Specifically, an optimization problem is solved to find the optimal operators $A$ and $B$, i.e., the operators that ensure that the resulting DD-ROM trajectories are as close as possible (typically in a least-squares sense) to the available data.
DD-ROM examples include the dynamic mode decomposition (DMD)~\cite{kutz2016dynamic,rowley2009spectral,schmid2010dynamic},
Koopman theory~\cite{mezic2005spectral},
the sparse identification of nonlinear dynamics (SINDy) algorithm~\cite{brunton2016discovering}, and 
the operator inference method~\cite{peherstorfer2015dynamic,peherstorfer2016data}.

\bigskip

We propose a {\it hybrid projection/data-driven ROM (Proj-DD-ROM)}~\cite{buffoni2006low,chekroun2017closures,couplet2005calibrated,galletti2007accurate,galletti2004low,gouasmi2016characterizing,lu2017data,noack2005need} in order to combine the best parts of each approach. 
We use the projection to determine the operators $A$ and $B$ in~\eqref{eqn:proj-rom-tau} and data-driven modeling to determine the unknown $\btau$ in~\eqref{eqn:proj-rom-tau}, which models the interaction between resolved  and unresolved modes.
The resulting ROM, which we call the {\it data-driven filtered ROM (DDF-ROM)}, is schematically illustrated in~\eqref{eqn:ddf-rom-schematic}, below.
\begin{equation}
\boxed{
\begin{CD}
	\text{FOM}		
	@>
	\text{filtering + projection} 
	>>	
	\text{F-ROM}
	@>
	\text{data-driven modeling} 
	>>	
	\text{DDF-ROM} \\[0.1cm]
\end{CD}
}
	\label{eqn:ddf-rom-schematic}
\end{equation}

In the first step of the DDF-ROM~\eqref{eqn:ddf-rom-schematic}, we use a projection approach to find the operators $A$ and $B$ in~\eqref{eqn:proj-rom-tau} as well as an {\it explicit formula} for $\btau$. 
To this end, we use a {\it ROM spatial filter} to filter the FOM, which contains all the information in the underlying system.
The resulting {\it filtered ROM (F-ROM)} approximates only the large spatial structures of the system and, therefore, requires fewer modes than the FOM.
The F-ROM takes the following form:
\begin{eqnarray}
	\dot{\ba}
	= A \, \ba
	+ \ba^{\top} \, B \, \ba
	+ \btau(FOM) \, ,
	\label{eqn:filtered-rom}
\end{eqnarray}
where $\btau(FOM)$ denotes the explicit dependence of $\btau$ on the FOM data.

We emphasize that to make the F-ROM~\eqref{eqn:filtered-rom} usable, one still needs to solve the {\it ROM closure problem}, i.e., to determine a formula of the form 
\begin{eqnarray}
	\btau(FOM)
	\approx \btau(\ba) \, .
	\label{eqn:f-rom-closure}
\end{eqnarray}
To this end, in the second step of the DDF-ROM~\eqref{eqn:ddf-rom-schematic}, we use {\it data-driven modeling}. 
First, we employ a quadratic ansatz to model $\btau$ in~\eqref{eqn:f-rom-closure}:
\begin{eqnarray}
	\btau(FOM)
	\approx \tA \, \ba
	+ \ba^{\top} \, \tB \, \ba \, .
	\label{eqn:f-rom-ansatz}
\end{eqnarray}
Then, we find $\tA$ and $\tB$ in~\eqref{eqn:f-rom-ansatz} by solving a (low-dimensional) optimization problem that minimizes the difference between $\btau$ calculated with the FOM data, and $\btau$ calculated with our ansatz:
\begin{eqnarray}
		\min_{\tA , \tB} \| \btau(FOM) - ( \tA \, \ba + \ba^{\top} \, \tB \, \ba) \|^2 \, .
	\label{eqn:f-rom-optimization}
\end{eqnarray}

At the end of the two steps of~\eqref{eqn:ddf-rom-schematic}, we obtain the DDF-ROM:
\begin{eqnarray}
\boxed{
	\dot{\ba}
	= (A + \tA) \, \ba
	+ \ba^{\top} \, (B + \tB) \, \ba
}
	\label{eqn:ddf-rom-intro}
\end{eqnarray}
  
\bigskip

We note that the hybrid Proj-DD-ROM approach used to construct the DDF-ROM~\eqref{eqn:ddf-rom-intro} is different from the DD-ROM approach.
Indeed, although we use data-driven modeling to develop the DDF-ROM, we do so to determine {\it only} $\tA$ and $\tB$ (and, thus, $\btau$).
This is in contrast with standard DD-ROMs, where data-driven modeling is used to build {\it all} the operators, i.e., not only $\tA$ and $\tB$, but also $A$ and $B$.

The DDF-ROM~\eqref{eqn:ddf-rom-intro} is also different from the classic Proj-ROMs, although the latter sometimes employ data-driven modeling.  
Indeed, Proj-ROMs generally use a dissipation ansatz (e.g., the eddy viscosity ansatz~\eqref{eqn:proj-rom-tau-ev}) to model $\btau$ in~\eqref{eqn:proj-rom-tau}.
Thus, available data can only be used to determine the tuning parameters of these dissipative mechanisms~\cite{benosman2017learning,protas2015optimal,wang2012proper}. 
In contrast, the DDF-ROM does not make any a priori assumptions regarding $\btau$ and data is used to determine {\it all} the components of $\btau$.
Thus, the DDF-ROM represents a {\it general} ROM framework that, in principle, can be used for the numerical simulation of any nonlinear PDE.
The key tool that allows us to use data-driven modeling to determine all the components of $\btau$ (as opposed to only the tuning parameters, as in Proj-ROMs) is the ROM spatial filtering, which yields an {\it explicit formula} for $\btau$.
Indeed, once we know what exactly we want to model, we can use available data to model it.

Finally, we note that the DDF-ROM framework has some connections to some other popular models, in particular the nonlinear Galerkin~\cite{FMRT01}, large eddy simulation (LES), and  variational multiscale~\cite{hughes1998variational} methods, since they all use the small-large scale separation. 
However, DDF-ROM is different from all these methods since it uses a data-driven modeling approach to approximate the interaction with the unresolved modes, whereas the other methods do not (see, however,~\cite{langford1999optimal}, for a notable exception).

\vspace*{0.3cm}

The rest of the paper is organized as follows:
In Section~\ref{sec:rom}, we present the standard Proj-ROM.
In Section~\ref{sec:f-rom}, we introduce the filtered ROM.
In Section~\ref{sec:ddf-rom}, we use data-driven modeling to solve the closure problem in the filtered ROM and to construct the DDF-ROM.
In Section~\ref{sec:numerical-results}, we investigate the DDF-ROM in the numerical simulation of a 2D channel flow past a circular cylinder.  
Finally, in Section~\ref{sec:conclusions}, we draw conclusions and outline future research directions.

\section{Projection Reduced Order Models (Proj-ROMs)}
    \label{sec:rom}

In this section, we briefly review the proper orthogonal decomposition (Section~\ref{sec:pod}) and the standard projection ROM (Section~\ref{sec:g-rom}).  
Although the new DDF-ROM framework can be applied to many types of nonlinear PDEs, to present the method it is necessary to pick a particular model, and so we select our favorite, the incompressible Navier-Stokes equations (NSE):
\begin{eqnarray}
    && \frac{\partial \bu}{\partial t}
    - Re^{-1} \Delta \bu
    + \bu \cdot \nabla \bu
    + \nabla p
    = {\bf 0} \, ,
    \label{eqn:nse-1}                                                         \\
    && \nabla \cdot \bu
    = 0 \, ,
    \label{eqn:nse-2}
\end{eqnarray}
where $\bu$ is the velocity, $p$ the pressure, and $Re$ the Reynolds number. 
We use the initial condition $\bu(\bx, 0) = \bu_0(\bx)$ and (for simplicity) homogeneous Dirichlet boundary conditions: $\bu(\bx, t) = {\bf 0}$.

    \subsection{Proper Orthogonal Decomposition (POD)}
        \label{sec:pod}
        One of the most popular reduced order modeling techniques is the
        proper orthogonal decomposition (POD)~\cite{HLB96,noack2011reduced,Sir87abc}. 
        For the snapshots
        $\{\bu^1_h,\ldots, \bu^{N_s}_h\}$, which are, e.g., finite element (FE) solutions of
        \eqref{eqn:nse-1}--\eqref{eqn:nse-2} at $N_s$ different time instances,
        the POD seeks a low-dimensional basis that approximates the snapshots
        optimally with respect to a certain norm. In this paper, we choose the commonly
        used $L^2$-norm. The solution of the minimization
        problem is equivalent to the solution of the eigenvalue problem
        $
            YY^TM_h \bphi_j = \lambda_j \bphi_j,
            \  j=1,\ldots,N,
        $
        where $\bphi_j$ and $\lambda_j$ denote the vector of the FE coefficients
        of the POD basis functions and the POD eigenvalues, respectively, $Y$
        denotes the snapshot matrix, whose columns correspond to the FE
        coefficients of the snapshots, $M_h$ denotes the FE mass matrix, and $N$
        is the dimension of the FE space $\bX^h$.
        The eigenvalues are real and non-negative, so they can be ordered as
        follows: $\lambda_1 \ge \lambda_2 \ge \ldots \ge \lambda_d \ge \lambda_{d + 1} = \ldots = \lambda_N = 0$, where $d$ is the rank of the snapshot matrix.
        The POD basis consists of the normalized functions $\{
        \bphi_{j}\}_{j=1}^{r}$, which correspond to the first $r\le N$ largest
        eigenvalues. Thus, the POD space is defined as $\bX^r := \text{span} \{
        \bphi_1, \ldots, \bphi_r \}$.

\subsection{Standard Galerkin ROM (G-ROM)}
    \label{sec:g-rom}

The POD approximation of the velocity is defined as 
\begin{equation}
	{\bu}_r({\bf x},t) 
	\equiv \sum_{j=1}^r a_j(t) \bphi_j({\bf x}) \, ,
	\label{eqn:g-rom-1}
\end{equation}
where $\{a_{j}(t)\}_{j=1}^{r}$ are the sought time-varying coefficients, which are determined by solving the following system of equations: $\forall \, i = 1, \ldots, r,$
    \begin{eqnarray}
        \left(
            \frac{\partial \bu_r}{\partial t} , \bphi_{i}
        \right)
        + Re^{-1} \, \left(
            \nabla \bu_r ,
            \nabla \bphi_{i}
        \right)
        + \biggl(
            (\bu_r \cdot \nabla) \, \bu_r ,
            \bphi_{i}
        \biggr)
        = 0 \, .
    \label{eqn:g-rom-weak}
    \end{eqnarray}
In~\eqref{eqn:g-rom-weak}, we assume that the modes $\{ \bphi_1, \ldots, \bphi_r \}$ are perpendicular to the discrete pressure space, which is the case if standard, conforming LBB stable elements (such as Taylor-Hood, Scott-Vogelius, the mini-element, etc.) are used for the snapshot creation.
Plugging~\eqref{eqn:g-rom-1} into~\eqref{eqn:g-rom-weak} yields the {\it Galerkin ROM (G-ROM)}:  
\begin{eqnarray}
	\dot{\ba}
	= A \, \ba
	+ \ba^{\top} \, B \, \ba \, ,
	\label{eqn:g-rom}
\end{eqnarray}
which can be written componentwise as follows: $\forall \, i = 1 \ldots r,$
\begin{equation}
	\dot{a}_i	
	 = \sum_{m=1}^{r} A_{i m} \, a_m(t)
	 + \sum_{m=1}^{r} \sum_{n=1}^{r} B_{i m n} \, a_n(t) \, a_m(t) \, ,
	\label{eqn:g-rom-2}
\end{equation}
where
\begin{eqnarray}
	A_{im}
	= - Re^{-1} \, \left( \nabla \bphi_m , \nabla \bphi_i \right),
	\qquad
	B_{imn}
	= - \bigl( \bphi_m \cdot \nabla \bphi_n , \bphi_i \bigr) \, .
	\label{eqn:g-rom-3b}
\end{eqnarray}

\section{Filtered ROMs (F-ROMs)}
	\label{sec:f-rom}
	
In this section, we present details regarding the first step of the DDF-ROM framework~\eqref{eqn:ddf-rom-schematic}.  
That is, we discuss the creation of the {\it filtered ROM (F-ROM)}, from which the ROM closure problem will reveal itself.  
To this end, in Section~\ref{sec:explicit-rom-spatial-filtering} we present the ROM projection, which is the explicit ROM spatial filter that we use to construct the DDF-ROM.
In Section~\ref{sec:f-rom-framework}, we develop the F-ROM framework, which has also been used to develop LES-ROMs~\cite{xie2017approximate}.
Finally, in Section~\ref{sec:rom-closure-problem}, we outline the celebrated ROM closure problem, which needs to be solved in the F-ROM.
We emphasize that {\it the ROM closure problem is treated completely differently in DDF-ROM and LES-ROM}: DDF-ROM uses data-driven modeling, while LES-ROMs generally use phenomenological arguments (e.g., energy cascade and eddy viscosity). 

\subsection{ROM Spatial Filtering}
    \label{sec:explicit-rom-spatial-filtering}

Spatial filtering has been used in ROMs, mainly as a preprocessing tool to eliminate the noise in the snapshot data (see, e.g., Section 5 in~\cite{aradag2011filtered} for a survey of relevant work).  
However, our approach is fundamentally different:
We explicitly use spatial filtering in the construction of the actual ROM, not in the development of the ROM basis.  
In this paper, we exclusively use the ROM projection~\cite{wang2012proper,wells2017evolve} as a   spatial filter, but we note that we could also use other spatial filters (e.g., the ROM differential filter~\cite{wells2017evolve,xie2017approximate}).

For a fixed $r \leq d$ and a given $\bu \in \bX^h$, the ROM projection~\cite{wang2012proper,wells2017evolve} seeks $\obu^r \in \bX^{r}$ such that
        \begin{eqnarray}
            \left( \obu^r, \bphi_j \right)
            = ( \bu , \bphi_j )
            \quad \forall \, j=1, \ldots r \, .
            \label{eqn:rom-projection}
        \end{eqnarray}

\subsection{F-ROM Framework}
\label{sec:f-rom-framework}

To outline the F-ROM framework, we use the standard LES approach~\cite{layton2012approximate,rebollo2014mathematical,Sag06}, which consists of the following steps:
(i) Use an explicit spatial filter to filter the NSE.
(ii) Use the resulting spatially filtered NSE and the ROM approximation to obtain the F-ROM.

Filtering the NSE, assuming that differentiation and filtering commute~\cite{Sag06}, and projecting the resulting equations onto a space of weakly divergence-free functions $\boldsymbol{\phi}$, we obtain the spatially filtered NSE (see equations (35)--(36) in~\cite{xie2017approximate}):
\begin{eqnarray}
	\left(
            \frac{\partial \obu}{\partial t} , \boldsymbol{\phi}
        \right)
        + Re^{-1} \, \biggl(
            \nabla \obu ,
            \nabla \boldsymbol{\phi}
        \biggr)
        + \biggl(
            \bigl( \obu \cdot \nabla \bigr) \, \obu  ,
            \boldsymbol{\phi}
        \biggr)
        + \biggl(
            \btau^{SFS} ,
            \boldsymbol{\phi}
        \biggr)
        = {\bf 0} \, ,
    \label{eqn:les-rom-1}
\end{eqnarray}
where
\begin{eqnarray}
\btau^{SFS}
= \overline{ \bigl( \bu \cdot \nabla \bigr) \, \bu}
- \bigl( \obu \cdot \nabla \bigr) \, \obu
\label{eqn:les-rom-2}
\end{eqnarray}
is the {\it subfilter-scale stress tensor}.
    
The spatial structures in the spatially filtered NSE~\eqref{eqn:les-rom-1} are larger than the spatial structures in the NSE~\eqref{eqn:nse-1}.
Thus, we expect that for a fixed target numerical accuracy of the ROM, the spatially filtered NSE will require fewer POD modes than the NSE, which is advantageous from a computational point of view.

Of course, to develop a useful ROM from the spatially filtered NSE~\eqref{eqn:les-rom-1}, we need to (i) use a ROM approximation for the continuous velocity field $\obu$, and (ii) use a ROM approximation of the spatial filter.
We use $\bu_d \sim \bu$ in (i) (where $d$ is the rank of the snapshot matrix) and $\obu^r \sim \obu$ in (ii) (where $\obu^r$ is the ROM projection~\eqref{eqn:rom-projection}).

Using $\bu_d \sim \bu$ in (i) means that we employ the best possible approximation of the continuous velocity field $\bu$ in the set of snapshots (i.e., in $\bX^d$). 
Using the ROM projection onto $\bX^r$ as the spatial filter in (ii) means that we are projecting the equations from $\bX^d$ onto $\bX^r$.
With these choices in the spatially filtered NSE~\eqref{eqn:les-rom-1}, we obtain: 
$\forall \, i = 1, \ldots, r,$
\begin{eqnarray}
	\left(
            \frac{\partial \overline{\bu_d}^r}{\partial t} , \bphi_i
        \right)
        + Re^{-1} \, \biggl(
            \nabla  \overline{\bu_d}^r ,
            \nabla \bphi_i
        \biggr)
        + \biggl(
            \bigl( \overline{\bu_d}^r \cdot \nabla \bigr) \, \overline{\bu_d}^r ,
            \bphi_i
        \biggr)
        + \biggl(
            \btau_r^{SFS} ,
            \bphi_i
        \biggr)
        = {\bf 0} \, ,
    \label{eqn:f-rom-0}
\end{eqnarray}
where
\begin{eqnarray}
\btau_r^{SFS}
= \overline{\bigl( {\bu_d} \cdot \nabla \bigr) \, {\bu_d}}^r 
- \bigl( \overline{\bu_d}^r \cdot \nabla \bigr) \, \overline{\bu_d}^r \, .
\label{eqn:les-rom-4}
\end{eqnarray}
Since we are using the ROM projection onto $\bX^r$ as the spatial filter, we have
\begin{eqnarray}
\overline{\bu_d}^r
= \bu_r .
\label{eqn:les-rom-4b}
\end{eqnarray}
Plugging~\eqref{eqn:les-rom-4b} in~\eqref{eqn:f-rom-0}--\eqref{eqn:les-rom-4}, we obtain the following system of equations:
\begin{eqnarray}
	\left(
            \frac{\partial \bu_r}{\partial t} , \bphi_i
        \right)
        + Re^{-1} \, \biggl(
            \nabla \bu_r ,
            \nabla \bphi_i
        \biggr)
        + \biggl(
            \bigl( \bu_r \cdot \nabla \bigr) \, \bu_r ,
            \bphi_i
        \biggr)
        + \biggl(
            \btau_r^{SFS} ,
            \bphi_i
        \biggr)
        = {\bf 0} \, ,
    \label{eqn:f-rom-weak}
\end{eqnarray}
where the ROM stress tensor is
\begin{eqnarray}
\btau_r^{SFS}
= \overline{\bigl( {\bu_d} \cdot \nabla \bigr) \, {\bu_d}}^r 
- \bigl( {\bu_r} \cdot \nabla \bigr) \, {\bu_r} \, .
\label{eqn:les-rom-6}
\end{eqnarray}
Now plugging~\eqref{eqn:g-rom-1} into~\eqref{eqn:f-rom-weak} yields the {\it F-ROM}:  
\begin{eqnarray}
	\dot{\ba}
	= A \, \ba
	+ \ba^{\top} \, B \, \ba 
	+ \btau \, ,
	\label{eqn:f-rom}
\end{eqnarray}
where $A$ and $B$ are given by~\eqref{eqn:g-rom-3b} and the components of $\btau$ are given by 
\begin{eqnarray}
	\tau_i
	= \biggl(
            \btau_r^{SFS} ,
            \bphi_i
        \biggr) \, ,
        \qquad 
        i = 1, \ldots, r \, .
\label{eqn:les-rom-6b}
\end{eqnarray}
\begin{remark}[F-ROM Consistency]
	We note that the F-ROM is consistent with the NSE.
	Indeed, the F-ROM~\eqref{eqn:f-rom} is just the projection of the full order model (i.e., the best representation of the NSE in the snapshot space, $\bX^d$) onto the ROM space, $\bX^r$.
	Thus, as $r \rightarrow d$, the F-ROM is expected to converge to the best representation of F-ROM in the snapshot space.
	
	We emphasize that many other ROMs (e.g., eddy viscosity ROMs~\cite{AHLS88,HLB96,wang2012proper}) are not consistent with the NSE. 
	Indeed, since the G-ROM is modified empirically, the resulting eddy viscosity ROM no longer corresponds to a Galerkin projection of the NSE~\cite{balajewicz2016minimal}. 
	\label{remark:f-rom-consistency}
\end{remark}

\subsection{F-ROM Closure Modeling}
    \label{sec:rom-closure-problem}

The F-ROM~\eqref{eqn:f-rom} is an $r$-dimensional ODE system for $\bur$.
Since $r \ll N$, the F-ROM~\eqref{eqn:f-rom} is a computationally efficient surrogate model for the FOM (i.e., the FE approximation of the NSE, which is an $N$-dimensional ODE system).
We emphasize, however, that the F-ROM~\eqref{eqn:f-rom} is not a closed system of equations, since the ROM stress tensor $\btau_r^{SFS}$ (which is used in the definition of $\btau$; see equation~\eqref{eqn:les-rom-6b}) depends on $\bu_d$ (see equation~\eqref{eqn:les-rom-6}).
Thus, to close the F-ROM~\eqref{eqn:f-rom}, we need to solve the ROM closure problem~\cite{chekroun2017closures,galletti2007accurate,lu2017data,noack2008finite,sirisup2004spectral,wang2012proper}, i.e., to find a formula $ \btau \approx \btau(\ba)$.  

Note that by neglecting the last term on the LHS of~\eqref{eqn:f-rom}, the F-ROM~\eqref{eqn:f-rom} is identical to the standard G-ROM~\eqref{eqn:g-rom}.
Thus, formally, one could write the following decomposition:
\begin{equation}
	\boxed{\text{F-ROM = G-ROM} + \btau}
	\label{eqn:f-rom-decomposition}
\end{equation}

\vspace*{0.1cm}

The decomposition~\eqref{eqn:f-rom-decomposition} is not new.
Indeed, for complex flows, the G-ROM is supplemented with extra terms, which generally provide a dissipation mechanism (e.g., eddy viscosity)~\cite{AHLS88,wang2012proper}.
However, what is new in the F-ROM is the {\it explicit formula for $\btau$} (see equations~\eqref{eqn:les-rom-6} and \eqref{eqn:les-rom-6b}), {\it which allows for the first time the use of data-driven modeling of the entire missing ROM information}.
We do exactly this in the next section.

\section{Data-Driven Filtered ROM (DDF-ROM)}
    \label{sec:ddf-rom}

In this section, we construct the new data-driven filtered ROM.
Specifically, we use data-driven modeling~\cite{brunton2016discovering,buffoni2006low,cacuci2013computational,cordier2010calibration,galletti2007accurate,galletti2004low,gouasmi2016characterizing,kutz2013data,lu2017data,noack2005need,peherstorfer2016data,wang2017physics} to solve the F-ROM closure problem, i.e., to find a formula $ \btau \approx \btau(\ba)$ in~\eqref{eqn:f-rom}.
To make the F-ROM~\eqref{eqn:f-rom} resemble the standard G-ROM~\eqref{eqn:g-rom}, we make the following ansatz: 
\begin{eqnarray}
     \btau(\ba)
	= \tA \, \ba
	+ \ba^{\top} \, \tB \, \ba 
 \, . 
	\label{eqn:c-rom-6}
\end{eqnarray}
Using ansatz~\eqref{eqn:c-rom-6} in the F-ROM~\eqref{eqn:f-rom} yields a closed system of equations.

To find $\tA$ and $\tB$ in~\eqref{eqn:c-rom-6}, we use data-driven modeling.
That is, we find $\tA$ and $\tB$ that ensure the highest accuracy of the vector $\btau$ in the F-ROM~\eqref{eqn:f-rom}.
To this end, we minimize the $L^2$-norm of the difference between $\btau$ computed with the FOM data and equations~\eqref{eqn:les-rom-6} and~\eqref{eqn:les-rom-6b}, and $\btau$ computed with the ansatz~\eqref{eqn:c-rom-6} and the ROM coefficients obtained from the snapshots, $\bas$.
The values $\bas(t_j)$, computed at snapshot time instances $t_j , \ j = 1, \ldots, M$, are obtained by projecting the corresponding snapshots $\bu(t_j) = \sum_{k=1}^{d} \as_k(t_j) \, \bphi_k$ onto the POD basis functions $\bphi_i$ and using the orthogonality of the POD basis functions: 
$\forall \, i = 1, \ldots, d, \ \forall \, j = 1, \ldots, M,$
\begin{equation}
	\as_i(t_j)
	= \biggl( \bu(t_j) , \bphi_i \biggr) \, .
	\label{eqn:c-rom-8}
\end{equation}

To find $\tA$ and $\tB$, we solve the following optimization problem~\cite{noack2005need,peherstorfer2016data}:
\begin{eqnarray}
	\min_{\substack{\tA \in \R^{r \times r} \\[0.1cm] \tB \in \R^{r \times r \times r}}} \, 
	\sum_{j = 1}^{M} \| \btau^{true}(t_j) - \btau^{ansatz}(t_j) \|^2
	\label{eqn:c-rom-11} \, ,
\end{eqnarray}
where $\| \cdot \|$ is the Euclidian norm in $\R^r$ and $\btau^{true}(t_j)$ is the {\it true} $\btau(t_j)$ computed from the snapshot data: $\forall \, i = 1, \ldots ,r, \ \forall \, j = 1, \ldots ,M,$
\begin{eqnarray}
	\tau_i^{true}(t_j)
	&=& \left( \overline{\bigl( \bu^{snap}_d(t_j) \cdot \nabla \bigr) \, \bu^{snap}_d(t_j)}^r 
	- \bigl( \bu^{snap}_r(t_j) \cdot \nabla \bigr) \, \bu^{snap}_r(t_j)  , \bphi_i \right)
	\nonumber \\
	&=& \left( \, 
				\overline{ 
							  \sum_{k_1=1}^{d} \, \sum_{k_2=1}^{d} a^{snap}_{k_1}(t_j) \, a^{snap}_{k_2}(t_j) \,  
							  \bigl( \bphi_{k_1} \cdot \nabla \bigr) \, \bphi_{k_2} 
							}^r 
		   \right.
	\nonumber \\
	&& -  \left. \sum_{k_3=1}^{r} \, \sum_{k_4=1}^{r} a^{snap}_{k_3}(t_j) \, a^{snap}_{k_4}(t_j) \, 
			\bigl( \bphi_{k_3}  \cdot \nabla \bigr)  \,  \bphi_{k_4}  , \bphi_i \right) \, ,
	\label{eqn:c-les-rom-3}
\end{eqnarray}
where
\begin{eqnarray}
	\bu^{snap}_d(t_j)
	= \sum_{k=1}^{d} a^{snap}_{k}(t_j) \, \bphi_{k} \, ,
	\qquad
	\bu^{snap}_r(t_j)
	= \sum_{k=1}^{r} a^{snap}_{k}(t_j) \, \bphi_{k} \, .
	\label{eqn:c-les-rom-3c}
\end{eqnarray}

\begin{remark}[Computational Efficiency]
	\label{remark:c-les-rom-2}
	In practical settings, where the rank of the snapshot matrix can be extremely large (e.g., $d= \mathcal{O}(1000)$), using $\bu^{snap}_d \in \bX^d$ in~\eqref{eqn:c-les-rom-3} would be very costly.
	One possible solution would be to replace $\bu^{snap}_d$ with, say, $\bu^{snap}_{2 r}$ and the ROM projection on $\bX^d$ with the ROM projection on $\bX^{2 r}$.
	We numerically investigate the accuracy of this approximation in Section~\ref{sec:numerical-results}.
\end{remark}

Finally, we compute $\btau^{ansatz}(t_j)$ from the ansatz~\eqref{eqn:c-rom-6} and the snapshot data:
\begin{equation}
	\btau^{ansatz}(t_j)
	= \tA \, \bas(t_j)
	+ \bas(t_j)^{\top} \, \tB \, \bas(t_j) \, . 
	\label{eqn:c-les-rom-4}
\end{equation}

Plugging~\eqref{eqn:c-rom-8},~\eqref{eqn:c-les-rom-3}, and~\eqref{eqn:c-les-rom-4} into the  minimization problem~\eqref{eqn:c-rom-11}, we obtain
\begin{equation}
		\min_{\substack{\tA \in \R^{r \times r} \\[0.1cm] \tB \in \R^{r \times r \times r}}} \, 
		\sum_{j = 1}^{M} \, \left\| \, \btau^{true}(t_j) - \tA \, \bas(t_j) - \bas(t_j)^{\top} \, \tB \, \bas(t_j) \, \right\|^2 \, ,
	\label{eqn:c-les-rom-12}
\end{equation}
where the vector $\btau^{true}(t_j) \in \R^{r \times 1}$ is defined in~\eqref{eqn:c-les-rom-3}.


Next, we introduce some notation that allows us to write the optimization problem~\eqref{eqn:c-les-rom-12} as a least squares problem.
To this end, we define the vector $\bx \in \R^{(r^2 + r^3) \times 1}$ that contains all the entries of $\tA$ and $\tB$ (i.e., the unknowns in the optimization problem~\eqref{eqn:c-les-rom-12}), and the vector $\bff \in \R^{(M \, r) \times 1}$ and matrix $E \in \R^{(M \, r) \times (r^2 + r^3)}$, which are computed from $\bas(t_j)$ and are chosen to satisfy the following equality~\cite{peherstorfer2016data}:
\begin{equation}
		\sum_{j = 1}^{M} \, \left\| \, \btau^{true}(t_j) - \tA \, \bas(t_j) - \bas(t_j)^{\top} \, \tB \, \bas(t_j) \, \right\|^2
		=  \left\| \, \bff - E \, \bx \, \right\|^2 \, .
	\label{eqn:c-les-rom-13}
\end{equation}
With this notation, the optimization problem~\eqref{eqn:c-les-rom-12} can be written as a {\it linear least squares} problem~\cite{peherstorfer2016data}:
\begin{equation}
		\min_{\substack{\bx \in \R^{(r^2 + r^3) \times 1}}} \, 
		\left\| \, \bff - E \, \bx \, \right\|^2 \, .
	\label{eqn:least-squares}
\end{equation}

The optimal $\tA^{opt}$ and $\tB^{opt}$ (i.e., the entries in $\bx$ that solves the linear least squares problem~\eqref{eqn:least-squares}) are used in the F-ROM~\eqref{eqn:f-rom}, yielding the {\it data-driven filtered ROM (DDF-ROM)}
\begin{equation}
	\boxed{
	\dot{\ba}
	= \left( A + \tA \right) \, \ba
	+ \ba^{\top} \, \left( B + \tB \right) \, \ba \, . 
	}
	\label{eqn:ddf-rom}
\end{equation}

In~\cite{peherstorfer2016data}, the authors note that data-driven least squares problems can be ill-conditioned.
To remedy this ill-conditioning, the authors use an empirical remedy: they combine trajectories of different initial conditions. 
In our numerical experiments in Section~\ref{sec:numerical-results}, the least squares problem~\eqref{eqn:least-squares} is also ill-conditioned, just as in~\cite{peherstorfer2016data}.
To tackle this challenge, however, we propose an approach that is different from that used in~\cite{peherstorfer2016data}:
We use the {\it truncated singular value decomposition (SVD)} (see Section 3.5 in~\cite{demmel1997applied}).

The DDF-ROM with the truncated SVD can be summarized in the following algorithm:

\begin{algorithm}[H]
	\caption{DDF-ROM}
	\label{alg:ddf-rom}
	\begin{algorithmic}[1]
		\STATE{
					Consider the F-ROM~\eqref{eqn:f-rom} 				
						\begin{eqnarray}
							\dot{\ba} 
							= A \, \ba 
							+ \ba^{\top} \, B \, \ba 
							+ \btau \, .
							\label{eqn:alg-ddf-rom-1}
						\end{eqnarray}
					}
		\STATE{
					Use snapshot data and~\eqref{eqn:c-les-rom-3} to compute the true vector $\btau$ in~\eqref{eqn:alg-ddf-rom-1}, $\btau^{true}$:
						\begin{eqnarray}
							\hspace*{-0.8cm}
							\tau_i^{true}(t_j)
							= \biggl( \, \overline{\bigl( \bu^{snap}_d(t_j) \cdot \nabla \bigr) \, \bu^{snap}_d(t_j)}^r 
							- \bigl( \bu^{snap}_r(t_j) \cdot \nabla \bigr) \, \bu^{snap}_r(t_j)  \, , \, \bphi_i \, \biggr)	.		
							\label{eqn:alg-ddf-rom-2}
						\end{eqnarray}
					}
		\STATE{
					Use snapshot data and~\eqref{eqn:c-rom-6} to define the ansatz vector $\btau$ in~\eqref{eqn:alg-ddf-rom-1}, $\btau^{ansatz}$:
						\begin{eqnarray}
							\btau^{ansatz}(t_j)
							= \tA \, \bas(t_j)
							+ \bas(t_j)^{\top} \, \tB \, \bas(t_j) \, . 
							\label{eqn:alg-ddf-rom-3}
						\end{eqnarray}
					}
		\STATE{
					Use all the entries of $\tA$ and $\tB$ in~\eqref{eqn:alg-ddf-rom-3} to define vector of unknowns, $\bx$.
					}
		\STATE{
					Use $\btau^{true}$ in~\eqref{eqn:alg-ddf-rom-2} and $\btau^{ansatz}$ in~\eqref{eqn:alg-ddf-rom-3} to assemble vector $\bff$ and matrix $E$ that satisfy~\eqref{eqn:c-les-rom-13}.
					}
		\STATE{
					Use the truncated SVD algorithm to solve the linear least squares problem~\eqref{eqn:least-squares}:
								\begin{eqnarray}
									\min_{\substack{\bx \in \R^{(r^2 + r^3) \times 1}}} \, 
									\left\| \, \bff - E \, \bx \, \right\|^2 \, .
									\label{eqn:alg-ddf-rom-4}
								\end{eqnarray}
					}
  			\begin{enumerate}
				\item[(i)] Calculate the SVD of $E$: 
					\begin{eqnarray}
						E = U \, \Sigma V^{\top} \, .
						\label{eqn:svd}
					\end{eqnarray}
				\item[(ii)] Specify tolerance $tol$.
				\item[(iii)] Construct matrix $\widehat{\Sigma}$ from $\Sigma$ as follows: 
							$\widehat{\sigma}_i = \sigma_i$ if $\sigma_i > tol$.
							(That is, keep only the entries in $\Sigma$ that are larger than $tol$.)
				\item[(iv)] Construct $\widehat{E}$, the truncated SVD of $E$:
					\begin{eqnarray}
						\widehat{E} = \widehat{U} \, \widehat{\Sigma} \, \widehat{V}^{\top} \, ,
						\label{eqn:truncated-svd}
					\end{eqnarray}
					where $\widehat{U}$ and $\widehat{V}$ are the entries of $U$ and $V$ in~\eqref{eqn:svd} that correspond to $\widehat{\Sigma}$.
				\item[(v)] The solution of the least squares problem~\eqref{eqn:alg-ddf-rom-4} is 
					\begin{eqnarray}
						\bx
						= \left( \widehat{V} \, \widehat{\Sigma}^{-1} \, \widehat{U}^{\top} \right) \, {\bf f} \, .
					\label{eqn:c-les-rom-25-ill-conditioned}
				\end{eqnarray}
			\end{enumerate}
		\STATE{
					The DDF-ROM has the following form:
								\begin{equation}
									\boxed{
									\dot{\ba}
									= \left( A + \tA \right) \, \ba
									+ \ba^{\top} \, \left( B + \tB \right) \, \ba
									} 
									\label{eqn:alg-ddf-rom-5}
								\end{equation}
								where $\tA$ and $\tB$ are the appropriate entries of $\bx$ found in~\eqref{eqn:c-les-rom-25-ill-conditioned}.
					}
\end{algorithmic}
\end{algorithm}

\section{Numerical Results}
	\label{sec:numerical-results}

In this section, we investigate the DDF-ROM~\eqref{eqn:ddf-rom} in the numerical simulation of a 2D channel flow past a circular cylinder at a Reynolds number $Re=100$.
This is a benchmark problem from \cite{ST96} that is often used for testing new methods; see, e.g.,~\cite{mohebujjaman2017energy}, which is the setting that we adopt here.

In Section~\ref{sec:test-problem-setup}, we describe the mathematical and computational setting of the test problem.
In Section~\ref{sec:snapshot-rom-generation}, we outline the snapshot and ROM generation.
In Section~\ref{sec:DDF-ROM-vs-g-rom}, we compare the DDF-ROM with the standard G-ROM, both in terms of numerical accuracy and computational efficiency.
In Section~\ref{sec:ddf-rom-parameter-sensitivity}, we perform a sensitivity study with respect to the DDF-ROM parameters.
Finally, in Section~\ref{sec:DDF-ROM-vs-state-of-the-art-roms}, we compare the DDF-ROM with some of the most recent ROMs, both in terms of numerical accuracy and computational efficiency.

\subsection{Test Problem Setup}
	\label{sec:test-problem-setup}

\begin{figure}[h!]
\begin{center}
\includegraphics[width=0.7\textwidth,height=0.24\textwidth, trim=0 0 0 0, clip]{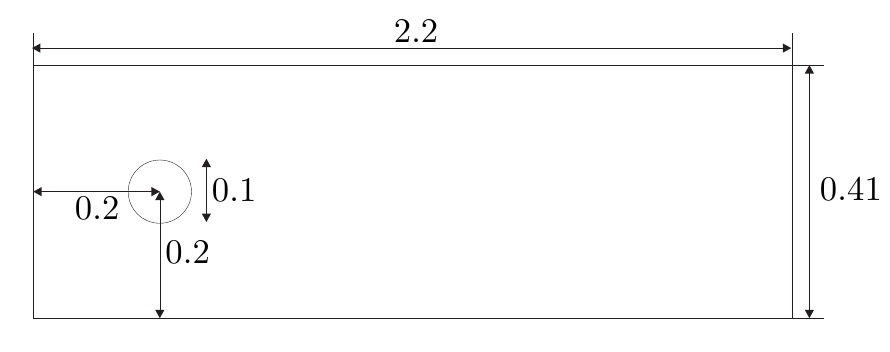}
\end{center}
\caption{\label{cyldomain} Channel flow around a cylinder domain.}
\end{figure}

The domain is a $2.2\times 0.41$ rectangular channel with a radius=$0.05$ cylinder, centered at $(0.2,0.2)$, see Figure~\ref{cyldomain}.  
No slip boundary conditions are prescribed for the walls and on the cylinder, and the inflow and outflow profiles are given by~\cite{john2004reference,rebholz2017improved} $u_{1}(0,y,t)=u_{1}(2.2,y,t)=\frac{6}{0.41^{2}}y(0.41-y) \, , u_{2}(0,y,t)=u_{2}(2.2,y,t)=0$.
The kinematic viscosity is $\nu=10^{-3}$, there is no forcing, and the flow starts from rest.

\subsection{Snapshot and ROM Generation}
	\label{sec:snapshot-rom-generation}

To compute the snapshots, we use the commonly used linearized BDF2 temporal discretization,  together with a FE spatial discretization utilizing the Scott-Vogelius element.  
On time step $1$, we use a backward Euler temporal discretization.  
All simulations use a time step size of $\Delta t=0.002$, are started from rest, and compute to the end time $T=17$.  
After an initial spin-up, the flow reaches a periodic-in-time (statistically steady) state by about $T=5$~\cite{mohebujjaman2017energy}.  
Snapshots are taken to be the solutions at each time step from $T=7$ to $T=7.332$, which corresponds to one period.
We compute on 3 different meshes, which provide approximately 103K, 35K, and 23K velocity degrees of freedom.  
The 103K mesh gives essentially a fully resolved solution, and the lift and drag predictions agree well with results from fine discretizations in~\cite{caiazzo2014numerical,ST96}:
$c_{d,max} = 3.2261,\ c_{l,max} = 1.0040.$
Results from the 35K meshes are only slightly less accurate, and error is more evident from the 23K  simulations.

\begin{figure}[h!]
	 	 	\centering
	 	 	\includegraphics[width = 1.2\textwidth, height=.3\textwidth,viewport=0 25 1000 330, clip]{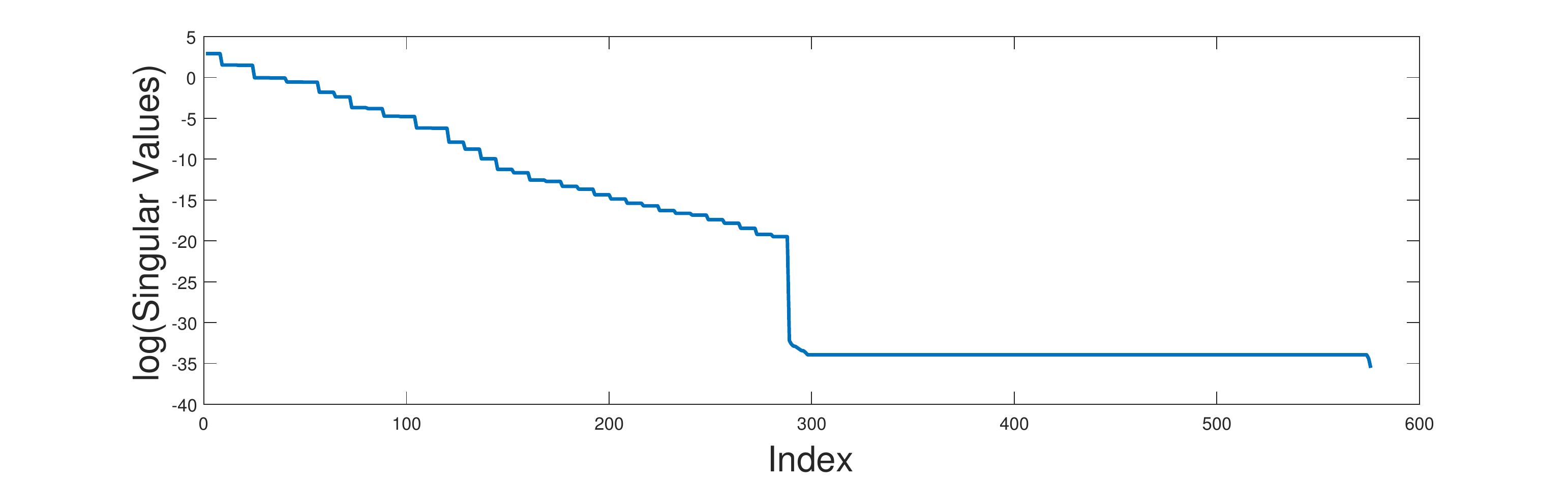}
	 	 	\caption{
			Plots of singular values  vs. index, for flow past a cylinder with $Re=100$.
			\label{fig:singular-values} 
			}
\end{figure}

The ROM modes are created from the snapshots in the usual way.  
The first mode is chosen to be the snapshot average, which satisfies the boundary conditions.  
This mode is then subtracted from the snapshots, and finally an eigenvalue problem is solved to find the dominant modes of these adjusted snapshots (see \cite{caiazzo2014numerical} for a more detailed description of the process). 
The singular values of the snapshot matrix are plotted in Figure~\ref{fig:singular-values}.

With the dominant modes created, the ROM is constructed as discussed in Section~\ref{sec:rom} using the BDF2 temporal discretization.  
In all of our tests, just as in the FE simulations, we take $\Delta t=0.002$; this choice creates no significant temporal error in any of our simulations (tests were done with varying $\Delta t$ to verify that 0.002 is sufficiently small).  
The ROM initial condition at $T=6.998$ is the $L^2$ projection of the FE solution at $T=6.998$ into the ROM space.
The ROM initial condition at $T=7$ is obtained by using the backward Euler method.
The ROMs are run from this initial time (now called $t=0$), and continued to $t=10$.
The ROMs are tested using three $r$ values: $r=8, r=10$, and $r=12$.
Lower $r$ values yield inaccurate results for all ROMs.

\subsection{DDF-ROM's Accuracy and Efficiency}
	\label{sec:DDF-ROM-vs-g-rom}

\paragraph{Accuracy}

In this section, we investigate the accuracy of the DDF-ROM.
To this end, we compare the DDF-ROM results with G-ROM results.
As benchmark, we use the results obtained on the finest mesh, which has 103K velocity degrees of freedom.  
To investigate the effect of the particular form of ansatz~\eqref{eqn:c-rom-6} on the numerical results, we consider two DDF-ROM versions: (i) DDF-ROM-quadratic, which is the standard DDF-ROM that considers both $\tA$ and $\tB$ in ansatz~\eqref{eqn:c-rom-6}, and (ii) DDF-ROM-linear, which is the  DDF-ROM that considers only $\tA$ in ansatz~\eqref{eqn:c-rom-6}.
Thus, in this section, we investigate three ROMs: DDF-ROM-quadratic, DDF-ROM-linear, and G-ROM.
We run all three ROMs with $r=8$ ROM modes.

\begin{figure}[h!]
			\centering
			\includegraphics[width = 1.2\textwidth, height=.3\textwidth,viewport=0 25 1000 260, clip]{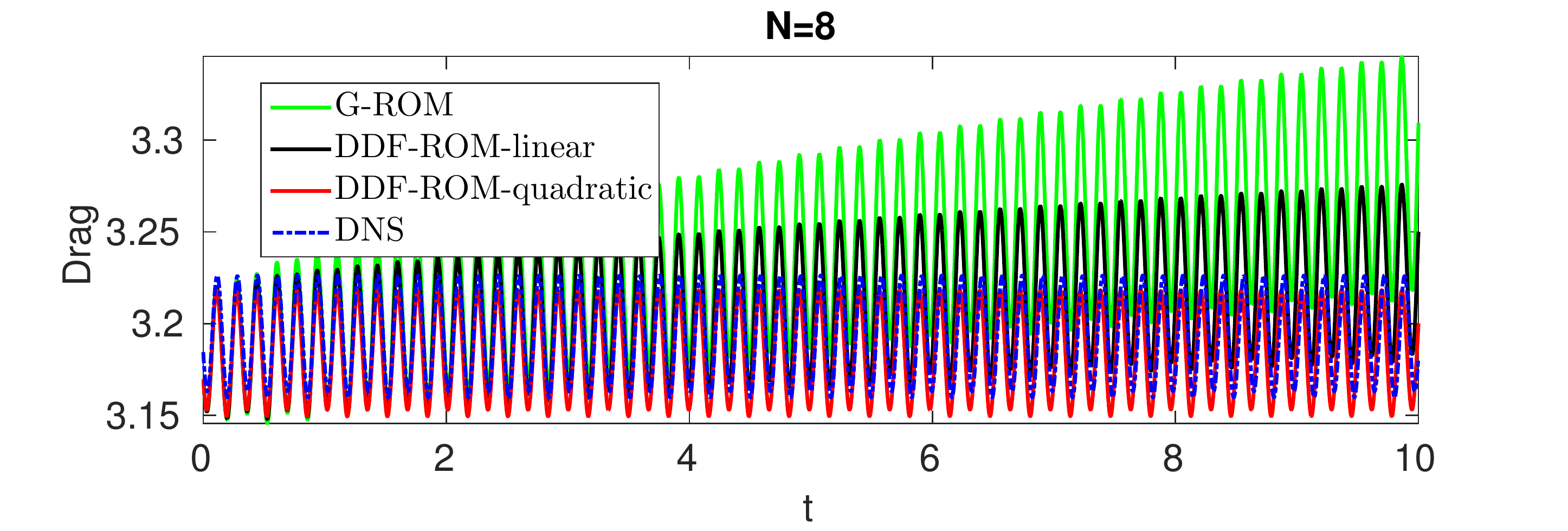}\\[0.5cm]
			\includegraphics[width = 1.2\textwidth, height=.3\textwidth,viewport=0 25 1000 270, clip]{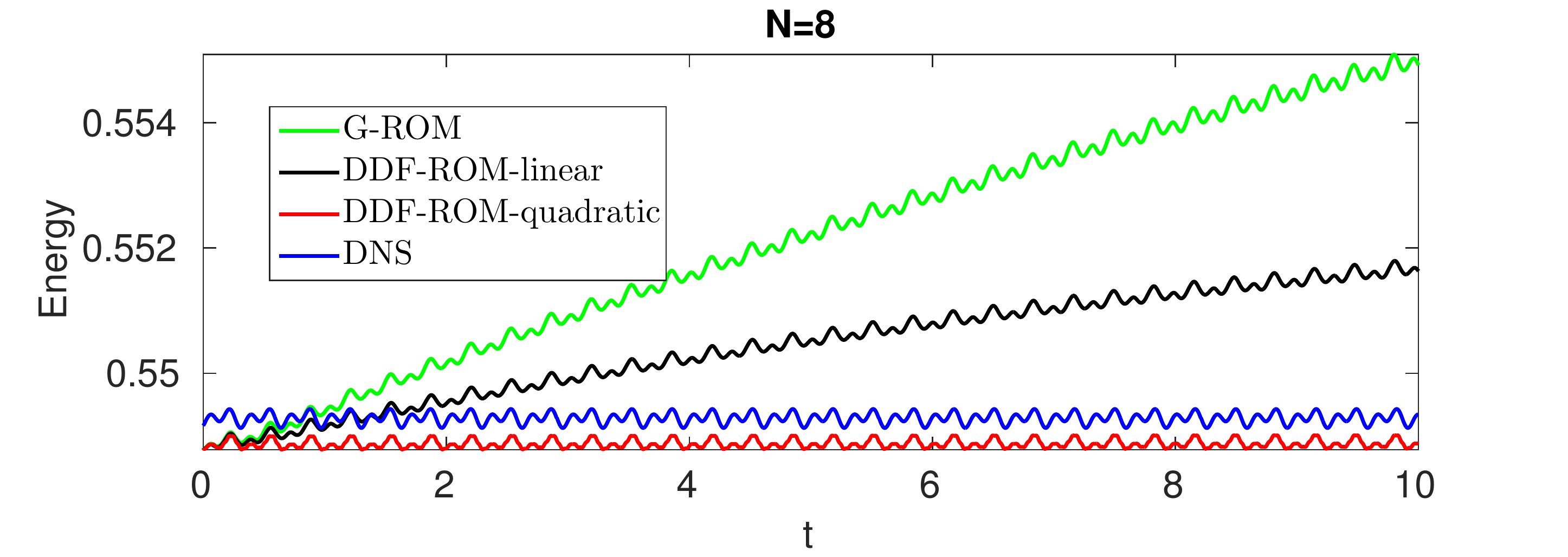}\\[0.5cm]
			\includegraphics[width = 1.2\textwidth, height=.3\textwidth,viewport=0 25 1000 260, clip]{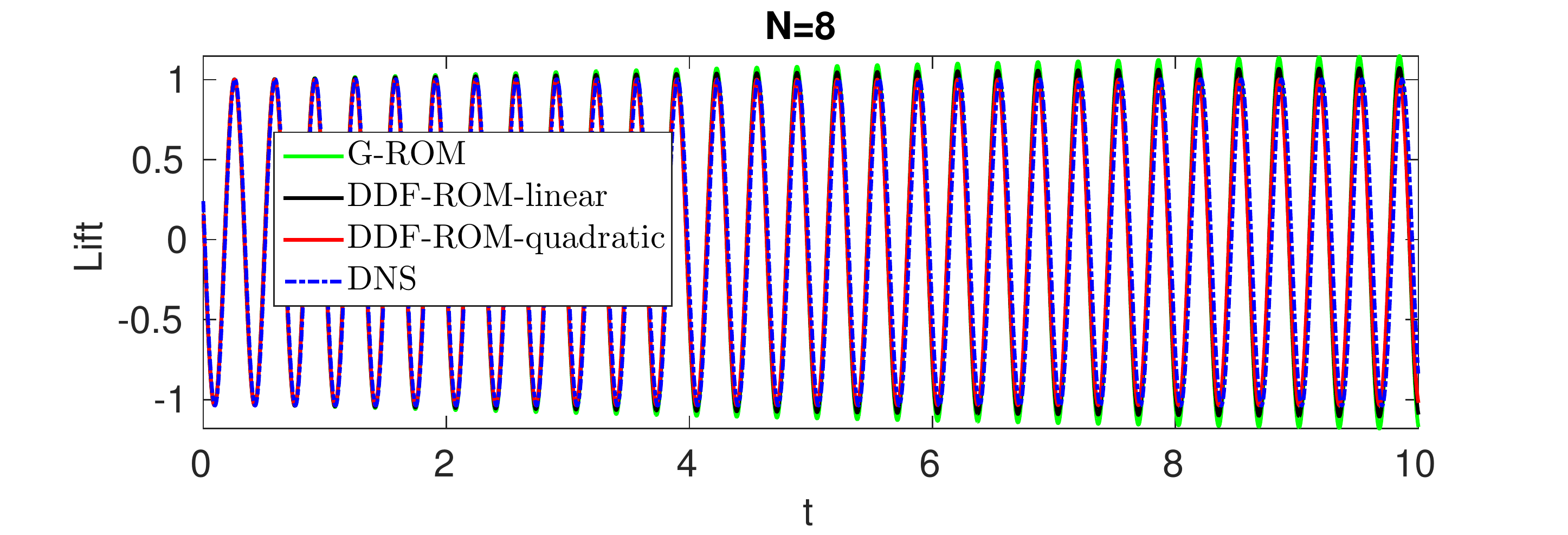}
	 	 	\caption{
			Plots of drag, energy, and lift coefficients vs. time for DDF-ROM-quadratic, DDF-ROM-linear, G-ROM, and DNS for flow past a cylinder with Re=100.
			\label{fig:drag-lift-energy}
			}
\end{figure}

In Figure~\ref{fig:drag-lift-energy}, we plot the drag, energy, and lift for all models. 
For both the DDF-ROM-quadratic and the DDF-ROM-linear, we use $tol=10^{-4}$ in the truncated SVD used in Step 6 of Algorithm~\ref{alg:ddf-rom}.
The main observation is that the DDF-ROM-quadratic performs the best.
This is especially true for the drag and energy plots.
The G-ROM performs the worst. 
The DDF-ROM-linear performs better than G-ROM, but worse than the DDF-ROM-quadratic.
In Table~\ref{table:c-rom-vs-g-rom}, we also list the average errors in the DDF-ROM-quadratic and the G-ROM (comparing to the direct numerical simulation (DNS) solution in the $L^2$-norm at each time step, and then taking the average over the time steps).  
For all $r$ values, the DDF-ROM-quadratic error is more than $50\%$ lower than the G-ROM error.
The plots in Figure~\ref{fig:drag-lift-energy} and the results in Table~\ref{table:c-rom-vs-g-rom} yield the following conclusions:
(i) the DDF-ROM-quadratic is significantly more accurate than the G-ROM, and 
(ii) the DDF-ROM-quadratic is more accurate than the DDF-ROM-linear.
Since we have shown that the DDF-ROM-quadratic is significantly more accurate than the DDF-ROM-linear, in what follows we only consider the DDF-ROM-quadratic.
Furthermore, since the DDF-ROM-quadratic is the standard DDF-ROM, we use the latter notation.

\paragraph{Efficiency}
Although the DDF-ROM is more accurate than the G-ROM, it is less efficient because it must calculate $\tA$ and $\tB$, and so a comparison is in order.  
We investigate the efficiency of the DDF-ROM and give offline timings for DDF-ROM and G-ROM in Table~\ref{table:c-rom-vs-g-rom}.
We note that the online timings for DDF-ROM and G-ROM were similar, so we did not include them in Table~\ref{table:c-rom-vs-g-rom}.
We also note that although the offline timings of both the DDF-ROM and the G-ROM are relatively large, they  could be significantly sped up with parallel computations.  

In its original form, the DDF-ROM uses~\eqref{eqn:c-les-rom-3} to compute $\btau^{true}(t_j)$.  
As mentioned in Remark~\ref{remark:c-les-rom-2}, this computation utilizes $\bu^{snap}_d \in \bX^d$.  
Since $d = \cO(1000)$ in some practical settings, this could make the DDF-ROM computationally costly.  
Hence, following Remark~\ref{remark:c-les-rom-2}, we replace $\bu^{snap}_d$ in~\eqref{eqn:c-les-rom-3} with $\bu^{snap}_{m}$, where $r \leq m \leq d$:
\begin{equation}
	\bu^{snap}_d
	\approx
	\bu^{snap}_{m}
	\qquad r \leq m \leq d \, .
	\label{eqn:numerical-results-4}
\end{equation}
When $m$ in~\eqref{eqn:numerical-results-4} is low (i.e., close to $r$), the DDF-ROM computational cost will be low, but the accuracy will also be low.
On the other hand, when $m$ in~\eqref{eqn:numerical-results-4} is large (i.e., close to $d$), the DDF-ROM accuracy will be high, but the computational cost will also be high.  
Thus, we need to find an $m$ value in~\eqref{eqn:numerical-results-4} that ensures a compromise between accuracy and efficiency in the DDF-ROM.  
From Table~\ref{table:c-rom-vs-g-rom}, we observe that $m=r+1$ seems to achieve compromise between accuracy and efficiency in the DDF-ROM.

\begin{table}[H]
\centering
\begin{tabular}{|l| l| c|c|c|c|c|c|c|}
\hline
Method &  Proj.  & \multicolumn{3}{c|}{ $ \| u_{DNS} - u_{ROM} \| $ } &  \multicolumn{3}{c|}{offline timing (seconds)} \\ \hline
                  &                  & r=8             & r=10       & r=12     	&	r=8           & r=10            & r=12        \\ \hline
G-ROM      &                  &  0.0159      & 0.0145   &  0.0039   &	855.52s    & 1567.44s    & 2708.83s  \\ \hline
DDF-ROM & $X^r$        &  0.0159       & 0.0144  & 0.0039 	&  1187.21s	  & 1902.04s    & 3043.01s \\ \hline
DDF-ROM & $X^{r+1}$  &  0.0099      &  0.0050  & 0.0020	&  1528.12s	  & 2433.48s    & 3753.97s \\ \hline
DDF-ROM & $X^{2r}$    &  0.0092      &  0.0050  & 0.0019	&  6373.97s	  & 12162.90s  & 21028.40s \\ \hline
DDF-ROM & $X^{3r}$    &  0.0092      &  0.0050  & 0.0019	&  21027.32s  & 40224.97s  & 69168.40s \\ \hline
\end{tabular}
\caption{
	DDF-ROM and G-ROM errors and offline timings for flow past a cylinder with Re=100, using varying $r$ and ROM projection spaces.
	\label{table:c-rom-vs-g-rom}	}
\end{table}

\subsection{DDF-ROM's Parameter Sensitivity}
	\label{sec:ddf-rom-parameter-sensitivity}

We also perform a sensitivity study on the DDF-ROM parameters.

In Table~\ref{tab:senstivity}, we list the DDF-ROM's lift and drag coefficients, and their ranges for different $r$ values ($r=8, r=10$, and $r=12$) and numbers of FE degrees of freedom (23K, 35K, and 103K).
We denote $C_d^{range} = |C_d^{max}-C_d^{min}|$ and $C_l^{range} = |C_l^{max}-C_l^{min}|$.
The results show that the DDF-ROM has a relatively low sensitivity with respect to $r$ and the number of FE degrees of freedom.

\begin{table}[h!]
\begin{center} 
  \begin{tabular}{|l|c|c||c||c||c||c||}
    \hline
  $r$ & FE dof  & $C_d^{ave}$ &  $C_d^{range}$ &  $C_l^{ave}$ & $C_l^{range}$ \\ \hline
        &   103K  & 3.16               &  0.07                  & -0.02             & 2.04                 \\ \hline \hline

  8  & 23K & 3.16 & 0.06  & -0.02 & 1.88 \\ \hline
  8  & 35K & 3.18 & 0.07  & -0.02 & 2.04 \\ \hline 
  8  & 103K & 3.19 & 0.07  & -0.02 & 2.05 \\ \hline  
  \hline
  10 & 23K  & 3.16 & 0.06 & -0.02 & 1.88 \\ \hline
  10 & 35K  & 3.18 & 0.07 & -0.02 & 2.04 \\ \hline
  10 & 103K & 3.19 & 0.07 & -0.02 & 2.04  \\ \hline
  \hline
  12 & 23K  & 3.16 & 0.06 & -0.02 & 1.88 \\ \hline
  12 & 35K  & 3.18 & 0.07 & -0.02 & 2.04 \\ \hline
 12 & 103K & 3.19 & 0.07 & -0.02 & 2.04 \\ \hline
  
    \end{tabular}
     \caption{
     	DDF-ROM average lift and drag coefficients, and their ranges for flow past a cylinder with Re=100, using different $r$ values and numbers of FE degrees of freedom (dof).
     	For comparison purposes, DNS results are listed in second row.
		\label{tab:senstivity} 
	}
\end{center}
\end{table}

\vspace*{-0.1cm}

We also perform a DDF-ROM sensitivity study with respect to changes in $tol$, which is the tolerance value used in the truncated SVD in Step 6 of Algorithm~\ref{alg:ddf-rom}.
Values around the value used to generate the plots in Figure~\ref{fig:drag-lift-energy} (i.e., $tol=10^{-4}$) yield similar results.
However, values that were significantly larger or lower than $tol=10^{-4}$ yield inaccurate results.
We conclude that the DDF-ROM results are sensitive with respect to $tol$.

\subsection{DDF-ROM vs. State-Of-The-Art ROMs}
	\label{sec:DDF-ROM-vs-state-of-the-art-roms}
	
Above we have shown that the new DDF-ROM is a clear improvement over the standard G-ROM, and a natural question is whether the DDF-ROM is also an improvement over other, more accurate ROMs.
To address this question, we consider three recently proposed ROMs for fluid flows:
two regularized ROMs (the Leray ROM (L-ROM) and the evolve-then-filter ROM (EF-ROM)~\cite{wells2017evolve}) and an LES-ROM (the approximate deconvolution ROM (AD-ROM)~\cite{xie2017approximate}).

\begin{figure}[h]
\centering
\includegraphics[scale=0.5]{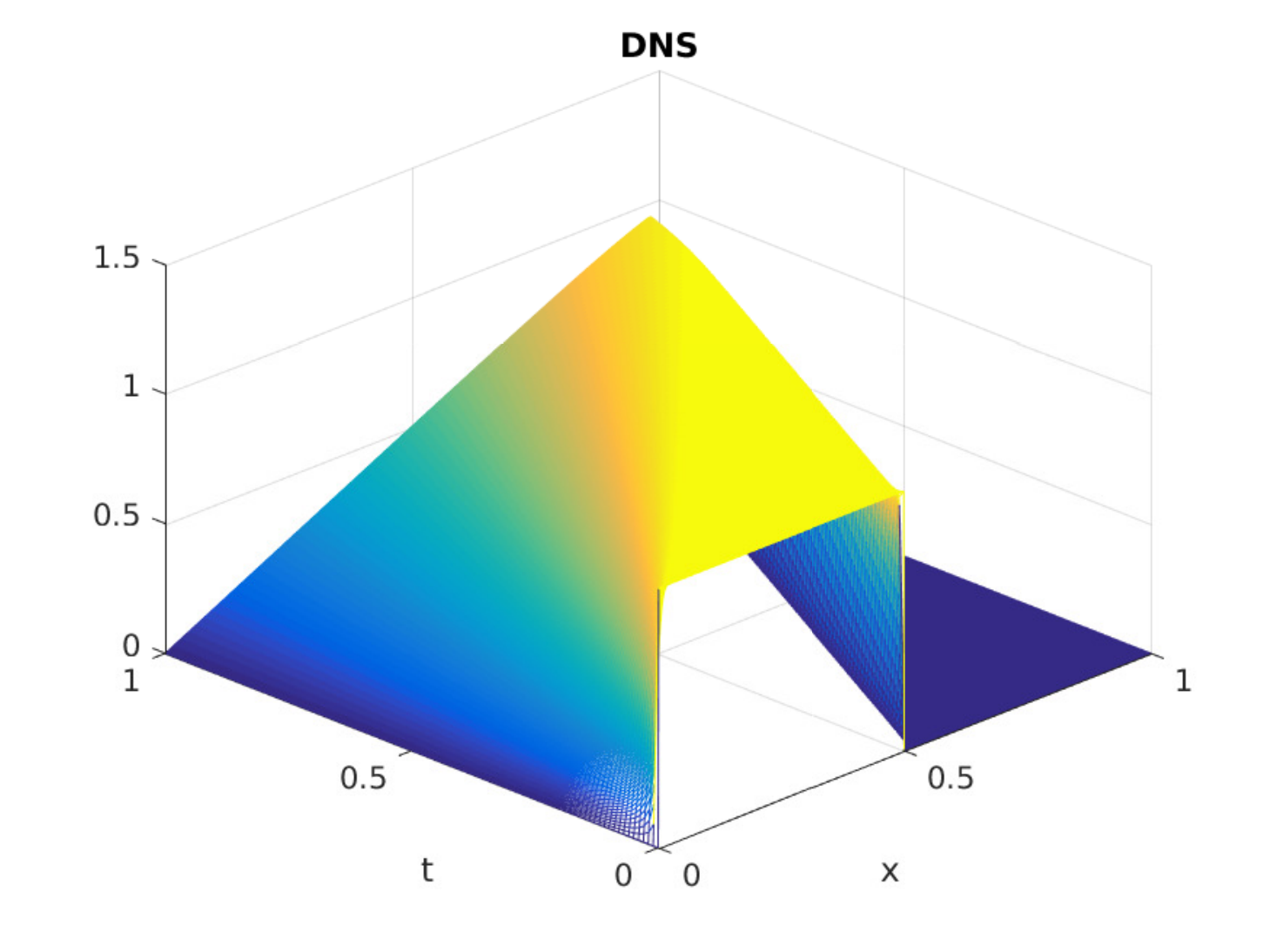}
\caption{Plot of the solution of the Burgers equation DNS.}
\label{fig:burgers-dns}
\end{figure}

We compare the new DDF-ROM with the AD-ROM, L-ROM, and EF-ROM.
We test all ROMs on the Burgers equation~\cite{KV01} with a steep internal layer (see Figure~\ref{fig:burgers-dns}) and with a small diffusion coefficient ($\nu=10^{-3}$).
We use the Burgers equation instead of the 2D flow past a circular cylinder that we utilize everywhere else in this section since AD-ROM, L-ROM, and EF-ROM results are available in the literature for the former, but not for the latter.

We list the errors in Table~\ref{tab:rom-comparison-error}. 
The DDF-ROM errors are slightly lower than all the other ROM errors.
We list the CPU times in Table~\ref{tab:rom-comparison-cpu-time}.
The DDF-ROM CPU time is significantly lower than the CPU times of the other ROMs.
The results in Table~\ref{tab:rom-comparison-error} and Table~\ref{tab:rom-comparison-cpu-time} consistently show that, for this test problem, the DDF-ROM is at least competitive with the other ROMs.
(These results are impressive, given that DDF-ROM-linear was used instead of the more accurate DDF-ROM-quadratic, see Section~\ref{sec:DDF-ROM-vs-g-rom}.)

\begin{table}[h]
\centering
\begin{tabular}{|l|c|c|c|c|}
\hline
       & L-DF & EF-ROM & AD-ROM & DDF-ROM \\ \hline
$r=6$  & 0.1385 & 0.1005 & 0.1096 & 0.0928 \\ \hline
$r=10$ & 0.1135 & 0.0699 & 0.0633 & 0.0627 \\ \hline
$r=15$ & 0.1037 & 0.0549 & 0.0532 & 0.0446 \\ \hline
\end{tabular}
\caption{
	Errors for L-ROM-DF, EF-ROM-DF, AD-ROM, and new DDF-ROM for Burgers equation.
	\label{tab:rom-comparison-error}
	}
\end{table}

\begin{table}[h]
\centering
\begin{tabular}{|l|c|c|c|c|}
\hline
       & L-DF & EF-ROM & AD-ROM & DDF-ROM \\ \hline
$r=6$  & 4.12 & 4.25 & 4.44 & 2.27 \\ \hline
$r=10$ & 6.72 & 6.91 & 7.26 & 4.42 \\ \hline
$r=15$ & 9.97 & 10.14 & 10.32 & 6.67 \\ \hline
\end{tabular}
\caption{
	CPU times for L-ROM-DF, EF-ROM-DF, AD-ROM, and new DDF-ROM for Burgers equation.
	\label{tab:rom-comparison-cpu-time}
	}
\end{table}

\vspace*{0.2cm} 

\section{Conclusions and Outlook}
	\label{sec:conclusions}

In this paper, we proposed a novel ROM framework for the numerical simulation of fluid flows.
This framework was based on explicit ROM spatial filtering and data-driven modeling.
The explicit ROM spatial filtering ensured computational efficiency of the filtered ROM, and the data-driven modeling was used to solve the ROM closure problem in the filtered ROM.

We numerically investigated the resulting DDF-ROM in the simulation of a 2D channel flow past a circular cylinder at a Reynolds number $Re=100$.
First, we compared the new DDF-ROM with the standard G-ROM.
The DDF-ROM was significantly more accurate than the G-ROM.
Furthermore, the computational costs of the DDF-ROM and G-ROM were similar, both costs being orders of magnitude lower than the computational cost of the full order model. 
For the 1D Burgers equation, we also compared the new DDF-ROM with state-of-the-art LES-ROMs.
The DDF-ROM was as accurate as state-of-the-art LES-ROMs.
However, the DDF-ROM was significantly more efficient than these LES-ROMs. 

Although these preliminary results are encouraging, the new DDF-ROM framework's full potential still needs to be explored.
Next, we outline several research directions that could be pursued. 

Probably the most important next step in the DDF-ROM development is the data-driven modeling used to solve the ROM closure problem in the filtered ROM.
In this paper, we have treated the entries in the subfilter-scale ROM stress tensor $\btau_r^{SFS}$ as general unknowns.
It is well known, however, that in fluid dynamics the subfilter-scale stress tensor satisfies important physical constraints~\cite{Sag06}.
We plan to replace the unconstrained optimization problem used in the data-driven modeling part of the DDF-ROM with a constrained optimization problem, which includes physical constraints for the subfilter-scale ROM stress tensor, such as energy conservation.
Similar approaches have been pursued in~\cite{kondrashov2015data,ling2016machine,loiseau2016constrained,majda2012physics} in different settings.

Another important research direction is the investigation of the generality of DDF-ROM. 
Although we constructed and tested the DDF-ROM in a fluid dynamics setting, the DDF-ROM framework can be applied to any type of nonlinear PDE that is amenable to reduced order modeling. 
Indeed, the only input needed in the DDF-ROM framework is the FOM data. 
Once those are supplied, the DDF-ROM proceeds in two steps.
(i) First, the given nonlinear PDE is spatially filtered.
The nonlinearity yields a nonlinear stress tensor (which will generally be different from the stress tensor $\btau_r^{SFS}$ used in this paper).
(ii) In the second step of the DDF-ROM construction, the available FOM data is used to compute an approximation for the true stress tensor in the filtered ROM in (i) and an optimization problem is solved to find the DDF-ROM coefficients.
We emphasize again that the entire DDF-ROM procedure does not use any phenomenological arguments that would restrict it to the particular physical system modeled by the given nonlinear PDE.
This is in stark contrast with, e.g., ROM closure models of eddy viscosity type~\cite{HLB96,osth2014need,wang2012proper}, which cannot be directly applied to other classes of PDEs.
Since the DDF-ROM is built upon general principles (i.e., filtering and data-driven modeling), we expect it to be successful in the numerical simulation of general mathematical models (e.g., from elasticity or bioengineering).

Finally, the ROM spatial filter used to build the DDF-ROM represents another research direction worthy of investigation.
In Section~\ref{sec:f-rom}, we used the ROM projection~\eqref{eqn:rom-projection} as a ROM spatial filter to construct the F-ROM~\eqref{eqn:f-rom}.
This choice of ROM spatial filter allowed us to write the F-ROM decomposition in~\eqref{eqn:f-rom-decomposition} and to explain why most ROM closure models amount to adding extra terms to the standard G-ROM.
We emphasize, however, that other ROM spatial filters could be used in the new DDF-ROM framework.
For example, the ROM differential filter (which was successfully used in developing LES-ROMs~\cite{wells2017evolve,xie2017approximate}) could be used as a ROM spatial filter to construct the F-ROM~\eqref{eqn:f-rom}.

\bibliographystyle{siamplain}
\bibliography{traian-siam}

\begin{thebibliography}{10}

\bibitem{amsallem2012stabilization}
{\sc D.~Amsallem and C.~Farhat}, {\em Stabilization of projection-based
  reduced-order models}, Int. J. Num. Meth. Eng., 91 (2012), pp.~358--377.

\bibitem{antoulas2005approximation}
{\sc A.~C. Antoulas}, {\em Approximation of large-scale dynamical systems},
  vol.~6, SIAM, 2005.

\bibitem{aradag2011filtered}
{\sc S.~Aradag, S.~Siegel, J.~Seidel, K.~Cohen, and T.~McLaughlin}, {\em
  Filtered {POD}-based low-dimensional modeling of the {3D} turbulent flow
  behind a circular cylinder}, Int. J. Num. Meth. Fluids, 66 (2011), pp.~1--16.

\bibitem{AHLS88}
{\sc N.~Aubry, P.~Holmes, J.~L. Lumley, and E.~Stone}, {\em The dynamics of
  coherent structures in the wall region of a turbulent boundary layer}, J.
  Fluid Mech., 192 (1988), pp.~115--173.

\bibitem{balajewicz2013low}
{\sc M.~J. Balajewicz, E.~H. Dowell, and B.~R. Noack}, {\em Low-dimensional
  modelling of high-{R}eynolds-number shear flows incorporating constraints
  from the {N}avier--{S}tokes equation}, J. Fluid Mech., 729 (2013),
  pp.~285--308.

\bibitem{balajewicz2016minimal}
{\sc M.~J. Balajewicz, I.~Tezaur, and E.~H. Dowell}, {\em {Minimal subspace
  rotation on the Stiefel manifold for stabilization and enhancement of
  projection-based reduced order models for the compressible Navier--Stokes
  equations}}, J. Comput. Phys., 321 (2016), pp.~224--241.

\bibitem{ballarin2016fast}
{\sc F.~Ballarin, E.~Faggiano, S.~Ippolito, A.~Manzoni, A.~Quarteroni,
  G.~Rozza, and R.~Scrofani}, {\em Fast simulations of patient-specific
  haemodynamics of coronary artery bypass grafts based on a {POD--Galerkin}
  method and a vascular shape parametrization}, J. Comput. Phys., 315 (2016),
  pp.~609--628.

\bibitem{ballarin2015supremizer}
{\sc F.~Ballarin, A.~Manzoni, A.~Quarteroni, and G.~Rozza}, {\em Supremizer
  stabilization of {POD--G}alerkin approximation of parametrized steady
  incompressible {N}avier--{S}tokes equations}, Int. J. Numer. Meth. Engng.,
  102 (2015), pp.~1136--1161.

\bibitem{barone2009stable}
{\sc M.~F. Barone, I.~Kalashnikova, D.~J. Segalman, and H.~K. Thornquist}, {\em
  Stable {G}alerkin reduced order models for linearized compressible flow}, J.
  Comput. Phys., 228 (2009), pp.~1932--1946.

\bibitem{benner2015survey}
{\sc P.~Benner, S.~Gugercin, and K.~Willcox}, {\em A survey of projection-based
  model reduction methods for parametric dynamical systems}, SIAM Rev., 57
  (2015), pp.~483--531.

\bibitem{benosman2017learning}
{\sc M.~Benosman, J.~Borggaard, O.~San, and B.~Kramer}, {\em {Learning-based
  robust stabilization for reduced-order models of 2D and 3D Boussinesq
  equations}}, Appl. Math. Model., 49 (2017), pp.~162--181.

\bibitem{bergmann2009enablers}
{\sc M.~Bergmann, C.~H. Bruneau, and A.~Iollo}, {\em {Enablers for robust POD
  models}}, J. Comput. Phys., 228 (2009), pp.~516--538.

\bibitem{bistrian2015improved}
{\sc D.~A. Bistrian and I.~M. Navon}, {\em An improved algorithm for the
  shallow water equations model reduction: {D}ynamic mode decomposition vs
  {POD}}, Int. J. Num. Meth. Fluids, 78 (2015), pp.~552--580.

\bibitem{brunton2016discovering}
{\sc S.~L. Brunton, J.~L. Proctor, and J.~N. Kutz}, {\em Discovering governing
  equations from data by sparse identification of nonlinear dynamical systems},
  Proc. Natl. Acad. Sci., 113 (2016), pp.~3932--3937.

\bibitem{buffoni2006low}
{\sc M.~Buffoni, S.~Camarri, A.~Iollo, and M.~V. Salvetti}, {\em
  {Low-dimensional modelling of a confined three-dimensional wake flow}}, J.
  Fluid Mech., 569 (2006), pp.~141--150.

\bibitem{cacuci2013computational}
{\sc D.~G. Cacuci, I.~M. Navon, and M.~Ionescu-Bujor}, {\em Computational
  methods for data evaluation and assimilation}, CRC Press, 2013.

\bibitem{caiazzo2014numerical}
{\sc A.~Caiazzo, T.~Iliescu, V.~John, and S.~Schyschlowa}, {\em A numerical
  investigation of velocity-pressure reduced order models for incompressible
  flows}, J. Comput. Phys., 259 (2014), pp.~598--616.

\bibitem{carlberg2017galerkin}
{\sc K.~Carlberg, M.~Barone, and H.~Antil}, {\em {Galerkin v. least-squares
  Petrov--Galerkin projection in nonlinear model reduction}}, J. Comput. Phys.,
  330 (2017), pp.~693--734.

\bibitem{chekroun2017closures}
{\sc M.~D. Chekroun, H.~Liu, J.~C. McWilliams, and S.~Wang}, {\em Closures for
  stochastic partial differential equations driven by degenerate noise}, In
  preparation,  (2017).

\bibitem{cordier2010calibration}
{\sc L.~Cordier, B.~Abou El~Majd, and J.~Favier}, {\em Calibration of {POD}
  reduced-order models using {T}ikhonov regularization}, Int. J. Num. Meth.
  Fluids, 63 (2010), pp.~269--296.

\bibitem{couplet2005calibrated}
{\sc M.~Couplet, C.~Basdevant, and P.~Sagaut}, {\em Calibrated reduced-order
  {POD}-{G}alerkin system for fluid flow modelling}, J. Comput. Phys., 207
  (2005), pp.~192--220.

\bibitem{CSB03}
{\sc M.~Couplet, P.~Sagaut, and C.~Basdevant}, {\em Intermodal energy transfers
  in a proper orthogonal decomposition--{G}alerkin representation of a
  turbulent separated flow}, J. Fluid Mech., 491 (2003), pp.~275--284.

\bibitem{demmel1997applied}
{\sc J.~W. Demmel}, {\em {Applied numerical linear algebra}}, Society for
  Industrial and Applied Mathematics Philadelphia, 1997.

\bibitem{FMRT01}
{\sc C.~Foia{\c{s}}, O.~Manley, R.~Rosa, and R.~Temam}, {\em Navier--{S}tokes
  {E}quations and {T}urbulence}, Cambridge University Press, 2001.

\bibitem{galletti2007accurate}
{\sc B.~Galletti, A.~Bottaro, C.-H. Bruneau, and A.~Iollo}, {\em Accurate model
  reduction of transient and forced wakes}, Eur. J. Mech. B-Fluid., 26 (2007),
  pp.~354--366.

\bibitem{galletti2004low}
{\sc B.~Galletti, C.~H. Bruneau, L.~Zannetti, and A.~Iollo}, {\em Low-order
  modelling of laminar flow regimes past a confined square cylinder}, J. Fluid
  Mech., 503 (2004), pp.~161--170.

\bibitem{giere2015supg}
{\sc S.~Giere, T.~Iliescu, V.~John, and D.~Wells}, {\em {SUPG} reduced order
  models for convection-dominated convection-diffusion-reaction equations},
  Comput. Methods Appl. Mech. Engrg., 289 (2015), pp.~454--474.

\bibitem{gouasmi2016characterizing}
{\sc A.~Gouasmi, E.~Parish, and K.~Duraisamy}, {\em Characterizing memory
  effects in coarse-grained nonlinear systems using the {M}ori-{Z}wanzig
  formalism}, 2016, \url{https://arxiv.org/abs/1611.06277}.

\bibitem{gunzburger2017ensemble}
{\sc M.~Gunzburger, N.~Jiang, and M.~Schneier}, {\em An ensemble-proper
  orthogonal decomposition method for the nonstationary {N}avier-{S}tokes
  equations}, SIAM J. Numer. Anal.,  (2017).
\newblock to appear.

\bibitem{hesthaven2015certified}
{\sc J.~S. Hesthaven, G.~Rozza, and B.~Stamm}, {\em Certified Reduced Basis
  Methods for Parametrized Partial Differential Equations}, Springer, 2015.

\bibitem{HLB96}
{\sc P.~Holmes, J.~L. Lumley, and G.~Berkooz}, {\em Turbulence, Coherent
  Structures, Dynamical Systems and Symmetry}, Cambridge, 1996.

\bibitem{hughes1998variational}
{\sc T.~J.~R. Hughes, G.~R. Feij{\'o}o, L.~Mazzei, and J.-B. Quincy}, {\em The
  variational multiscale method -- a paradigm for computational mechanics},
  Comput. Methods Appl. Mech. Engrg., 166 (1998), pp.~3--24.

\bibitem{john2004reference}
{\sc V.~John}, {\em Reference values for drag and lift of a two dimensional
  time-dependent flow around a cylinder}, Int. J. Num. Meth. Fluids, 44 (2004),
  pp.~777--788.

\bibitem{kondrashov2015data}
{\sc D.~Kondrashov, M.~D. Chekroun, and M.~Ghil}, {\em {Data-driven
  non-Markovian closure models}}, Phys. D, 297 (2015), pp.~33--55.

\bibitem{KV01}
{\sc K.~Kunisch and S.~Volkwein}, {\em Galerkin proper orthogonal decomposition
  methods for parabolic problems}, Numer. Math., 90 (2001), pp.~117--148.

\bibitem{kutz2013data}
{\sc J.~N. Kutz}, {\em Data-driven modeling \& scientific computation: methods
  for complex systems \& big data}, Oxford University Press, 2013.

\bibitem{kutz2016dynamic}
{\sc J.~N. Kutz, S.~L. Brunton, B.~W. Brunton, and J.~L. Proctor}, {\em Dynamic
  Mode Decomposition: Data-Driven Modeling of Complex Systems}, SIAM, 2016.

\bibitem{langford1999optimal}
{\sc J.~A. Langford and R.~D. Moser}, {\em {Optimal LES formulations for
  isotropic turbulence}}, J Fluid Mech., 398 (1999), pp.~321--346.

\bibitem{layton2012approximate}
{\sc W.~J. Layton and L.~G. Rebholz}, {\em Approximate Deconvolution Models of
  Turbulence: Analysis, Phenomenology and Numerical Analysis}, vol.~2042,
  Springer Berlin Heidelberg, 2012.

\bibitem{ling2016machine}
{\sc J.~Ling, R.~Jones, and J.~Templeton}, {\em Machine learning strategies for
  systems with invariance properties}, J. Comput. Phys., 318 (2016),
  pp.~22--35.

\bibitem{loiseau2016constrained}
{\sc J.-C. Loiseau and S.~L. Brunton}, {\em {Constrained sparse Galerkin
  regression}}, 2016, \url{https://arxiv.org/abs/1611.03271}.

\bibitem{lu2017data}
{\sc F.~Lu, K.~K. Lin, and A.~J. Chorin}, {\em {Data-based stochastic model
  reduction for the Kuramoto--Sivashinsky equation}}, Phys. D, 340 (2017),
  pp.~46--57.

\bibitem{majda2012physics}
{\sc A.~J. Majda and J.~Harlim}, {\em Physics constrained nonlinear regression
  models for time series}, Nonlinearity, 26 (2012), p.~201.

\bibitem{mezic2005spectral}
{\sc I.~Mezi{\'c}}, {\em Spectral properties of dynamical systems, model
  reduction and decompositions}, Nonlinear Dyn., 41 (2005), pp.~309--325.

\bibitem{mohebujjaman2017energy}
{\sc M.~Mohebujjaman, L.~G. Rebholz, X.~Xie, and T.~Iliescu}, {\em Energy
  balance and mass conservation in reduced order models of fluid flows}, J.
  Comput. Phys., 346 (2017), pp.~262--277.

\bibitem{noack2011reduced}
{\sc B.~R. Noack, M.~Morzynski, and G.~Tadmor}, {\em Reduced-Order Modelling
  for Flow Control}, vol.~528, Springer Verlag, 2011.

\bibitem{noack2005need}
{\sc B.~R. Noack, P.~Papas, and P.~A. Monkewitz}, {\em The need for a
  pressure-term representation in empirical {G}alerkin models of incompressible
  shear flows}, J. Fluid Mech., 523 (2005), pp.~339--365.

\bibitem{noack2008finite}
{\sc B.~R. Noack, M.~Schlegel, B.~Ahlborn, G.~Mutschke, M.~Morzynski, P.~Comte,
  and G.~Tadmor}, {\em {A finite-time thermodynamics of unsteady fluid flows}},
  J. Non-Equil. Thermody., 33 (2008), pp.~103--148.

\bibitem{osth2014need}
{\sc J.~{\"O}sth, B.~R. Noack, S.~Krajnovi{\'c}, D.~Barros, and J.~Bor{\'e}e},
  {\em On the need for a nonlinear subscale turbulence term in {POD} models as
  exemplified for a high-{R}eynolds-number flow over an {A}hmed body}, J. Fluid
  Mech., 747 (2014), pp.~518--544.

\bibitem{peherstorfer2015dynamic}
{\sc B.~Peherstorfer and K.~Willcox}, {\em Dynamic data-driven reduced-order
  models}, Comput. Methods Appl. Mech. Engrg., 291 (2015), pp.~21--41.

\bibitem{peherstorfer2016data}
{\sc B.~Peherstorfer and K.~Willcox}, {\em Data-driven operator inference for
  nonintrusive projection-based model reduction}, Comput. Methods Appl. Mech.
  Engrg., 306 (2016), pp.~196--215.

\bibitem{perotto2017higamod}
{\sc S.~Perotto, A.~Reali, P.~Rusconi, and A.~Veneziani}, {\em {HIGAMod: A
  Hierarchical IsoGeometric Approach for MODel reduction in curved pipes}},
  Comput. \& Fluids, 142 (2017), pp.~21--29.

\bibitem{protas2015optimal}
{\sc B.~Protas, B.~R. Noack, and J.~{\"O}sth}, {\em Optimal nonlinear eddy
  viscosity in {G}alerkin models of turbulent flows}, J. Fluid Mech., 766
  (2015), pp.~337--367.

\bibitem{quarteroni2015reduced}
{\sc A.~Quarteroni, A.~Manzoni, and F.~Negri}, {\em Reduced Basis Methods for
  Partial Differential Equations: An Introduction}, vol.~92, Springer, 2015.

\bibitem{rebholz2017improved}
{\sc L.~Rebholz and M.~Xiao}, {\em {Improved accuracy in algebraic splitting
  methods for Navier-Stokes equations}}, SIAM J. Sci. Comput.,  (2017).
\newblock to appear.

\bibitem{rebollo2014mathematical}
{\sc T.~C. Rebollo and R.~Lewandowski}, {\em Mathematical and Numerical
  Foundations of Turbulence Models and Applications}, Springer, 2014.

\bibitem{rowley2009spectral}
{\sc C.~W. Rowley, I.~Mezi{\'c}, S.~Bagheri, P.~Schlatter, and D.~S.
  Henningson}, {\em Spectral analysis of nonlinear flows}, J. Fluid Mech., 641
  (2009), pp.~115--127.

\bibitem{Sag06}
{\sc P.~Sagaut}, {\em Large Eddy Simulation for Incompressible Flows},
  Scientific Computation, Springer-Verlag, Berlin, third~ed., 2006.

\bibitem{ST96}
{\sc M.~Sch$\ddot{\mbox{a}}$fer and S.~Turek}, {\em The benchmark problem `flow
  around a cylinder' flow simulation with high performance computers {II}}, in
  E.H. Hirschel (Ed.), Notes on Numerical Fluid Mechanics, 52, Braunschweig,
  Vieweg (1996), pp.~547--566.

\bibitem{schmid2010dynamic}
{\sc P.~J. Schmid}, {\em Dynamic mode decomposition of numerical and
  experimental data}, J. Fluid Mech., 656 (2010), pp.~5--28.

\bibitem{sirisup2004spectral}
{\sc S.~Sirisup and G.~E. Karniadakis}, {\em {A spectral viscosity method for
  correcting the long-term behavior of POD models}}, J. Comput. Phys., 194
  (2004), pp.~92--116.

\bibitem{Sir87abc}
{\sc L.~Sirovich}, {\em Turbulence and the dynamics of coherent structures.
  {P}arts {I}--{III}}, Quart. Appl. Math., 45 (1987), pp.~561--590.

\bibitem{stefanescu2015pod}
{\sc R.~{\c{S}}tef{\u{a}}nescu, A.~Sandu, and I.~M. Navon}, {\em {POD/DEIM
  reduced-order strategies for efficient four dimensional variational data
  assimilation}}, J. Comput. Phys., 295 (2015), pp.~569--595.

\bibitem{wang2017physics}
{\sc J.-X. Wang, J.-L. Wu, and H.~Xiao}, {\em {Physics-informed machine
  learning approach for reconstructing Reynolds stress modeling discrepancies
  based on DNS data}}, Phys. Rev. Fluids, 2 (2017), p.~034603.

\bibitem{wang2012proper}
{\sc Z.~Wang, I.~Akhtar, J.~Borggaard, and T.~Iliescu}, {\em Proper orthogonal
  decomposition closure models for turbulent flows: A numerical comparison},
  Comput. Meth. Appl. Mech. Eng., 237-240 (2012), pp.~10--26.

\bibitem{wells2017evolve}
{\sc D.~Wells, Z.~Wang, X.~Xie, and T.~Iliescu}, {\em An evolve-then-filter
  regularized reduced order model for convection-dominated flows}, Int. J. Num.
  Meth. Fluids, 84 (2017), pp.~598--–615.

\bibitem{xie2017approximate}
{\sc X.~Xie, D.~Wells, Z.~Wang, and T.~Iliescu}, {\em Approximate deconvolution
  reduced order modeling}, Comput. Methods Appl. Mech. Engrg., 313 (2017),
  pp.~512--534.

\end{thebibliography}

\end{document}